\documentclass[10pt,conference]{IEEEtran}

\IEEEoverridecommandlockouts
\usepackage{cite}
\usepackage{amsmath,amssymb,amsfonts}
\usepackage{graphicx}
\usepackage{textcomp}
\usepackage{tabularx}
\usepackage{xcolor}
\usepackage{booktabs}

\def\BibTeX{{\rm B\kern-.05em{\sc i\kern-.025em b}\kern-.08em
    T\kern-.1667em\lower.7ex\hbox{E}\kern-.125emX}}
    
\usepackage{tabularx}
\usepackage{array,tabu}
\usepackage{siunitx}
\usepackage{multirow}
\usepackage[para]{threeparttablex}
\usepackage{qcircuit}
\usepackage{todonotes}
\usepackage{amsmath}
\usepackage{braket} % physics bracket shorthand
\usepackage{algorithm}
\usepackage{algpseudocode}
\floatname{algorithm}{Algorithm}

\usepackage{subcaption}
\usepackage{hyperref}

\newcommand{\intelphi}{Intel Xeon Phi}

\newcommand{\iphi}{Xeon Phi}

\begin{document}

% A command for highlighting updates for reviewers.
%\newcommand{\add}[1]{{\color{blue} #1}}
\newcommand{\add}[1]{{#1}}

\title{Tensor Network Quantum Simulator With Step-Dependent Parallelization}

 %  %\iffalse
 %  
 %  \author{Danylo Lykov}
 %  \email{dlykov@anl.gov}
 %  \affiliation{%
 %    \institution{Argonne National Laboratory}
 %    \streetaddress{9700 S. Cass Ave.}
 %    \city{Lemont} \state{IL} \postcode{60439} \country{USA}
 %  }
 %  
 %  \author{Roman Schutski}
 %  \email{r.schutski@skoltech.ru}
 %  \affiliation{%
 %    \institution{Rice University}
 %    \streetaddress{6100 Main St.}
 %    \city{Houston} \state{TX} \postcode{77005} \country{USA}
 %  }
 %  
 %  \author{Alexey Galda}
 %  \email{agalda@uchicago.edu}
 %  \affiliation{%
 %    \institution{University of Chicago}
 %    \institution{Argonne National Laboratory}
 %    \streetaddress{5801 S Ellis Ave.}
 %    \city{Chicago} \state{IL} \postcode{60637} \country{USA}
 %  }
 %  
 %  %\author{Roman Schutski}
 %  %\email{}
 %  %\affiliation{%
 %  %  \institution{Center for %Computational and Data-Intensive %Science and Engineering, %Skoltech, Skolkovo Innovation %Center}
 %  %  \streetaddress{Skoltech}
 %  %  \city{Moscow} %\postcode{121205} %\country{Russia}
 %  %}
 %  
 %  \author{Valerii Vinokur}
 %  \email{vinokour@anl.gov}
 %  \affiliation{%
 %    \institution{Argonne National Laboratory}
 %    \streetaddress{9700 S. Cass Ave.}
 %    \city{Lemont} \state{IL} \postcode{60439} \country{USA}
 %  }
 %  
 %  \author{Yuri Alexeev}
 %  \email{yuri@anl.gov}
 %  \affiliation{%
 %    \institution{Argonne National Laboratory}
 %    \streetaddress{9700 S. Cass Ave.}
 %    \city{Lemont} \state{IL} \postcode{60439} \country{USA}
 %  }

%\fi

\author{
    \IEEEauthorblockN{Danylo Lykov}
    \IEEEauthorblockA{Argonne National Laboratory\\9700 S. Cass Ave Lemont IL\\60439 USA}
    \and
    \IEEEauthorblockN{Roman Schutski}
    \IEEEauthorblockA{Rice University\\6100 Main St. Houston TX \\
    77005 USA}
    \and
    \IEEEauthorblockN{Alexey Galda}
    \IEEEauthorblockA{University of Chicago\\S Ellis Ave. Chicago IL \\60637 USA}
    \and
    \IEEEauthorblockN{Valeri Vinokur}
    \IEEEauthorblockA{Argonne National Laboratory\\9700 S. Cass Ave Lemont IL\\60439 USA}
    \and
    \IEEEauthorblockN{Yuri Alexeev}
    \IEEEauthorblockA{Argonne National Laboratory\\9700 S. Cass Ave Lemont IL\\60439 USA}
}
\newcommand{\stepslice}{\emph{step-dependent slicing} }

\maketitle

\begin{abstract}
In this work, we present a new large-scale quantum circuit simulator. 
It is based on the tensor network contraction technique to represent quantum circuits.
We propose a novel parallelization algorithm based on \stepslice .
In this paper, we push the requirement on the size of a quantum computer that will be needed to 
demonstrate the advantage of quantum computation with Quantum Approximate Optimization Algorithm (QAOA).
We computed a single amplitude of QAOA ansatz state on 210 qubits. The simulation involved 1,785 gates on 1,024 nodes of the the Cray XC 40 supercomputer Theta.
To the best of our knowledge, this constitutes the largest simulation of QAOA ansatz simulations reported to this date.

\end{abstract}

%\ccsdesc[500]{Theory of computation~Massively parallel algorithms}
%\ccsdesc[500]{Computing methodologies~Quantum mechanic simulation}
%\ccsdesc[500]{Computing methodologies~Massively parallel and high-performance simulations}


% xml stuff [...]
%% Keywords.  [...]

\begin{figure*}
\centering
\input{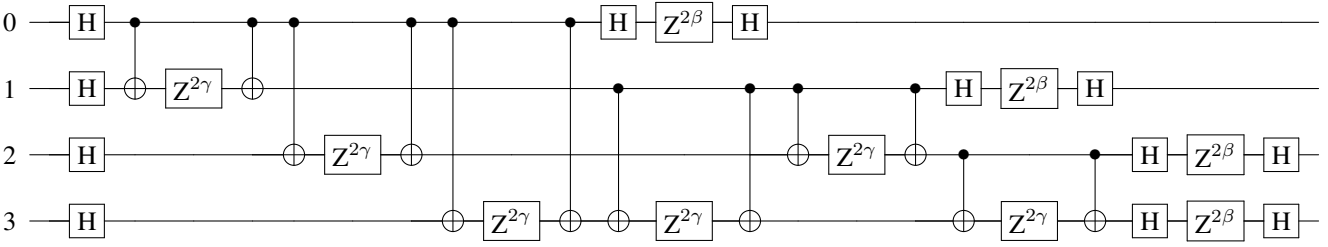}
\caption{p=1 depth QAOA circuit for a fully connected graph with 4 nodes.}
\label{fig:circ}
\end{figure*}
%GAIL - you do not refer to this figure in the text. 
\section{Introduction}

Simulations of quantum circuits on classical computers are essential for better understanding of how quantum computers operate, the optimization of their work, and the development of quantum algorithms. For example, simulators allow researchers to evaluate the complexity of new quantum algorithms and to develop and validate the design of new quantum circuits.

Many approaches have been proposed to simulate quantum circuits on classical computers. The major types of simulation techniques are full amplitude-vector evolution \cite{de2007massively, smelyanskiy2016qhipster, haner20170, wu2019full}, the Feynman paths approach \cite{bernstein1997quantum}, linear algebra open system simulation \cite{quac}, decision diagrams \cite{burgholzer_dd}, and tensor network contractions \cite{markov2008simulating, pednault2017breaking, boixo2017simulation}.

Tensor network contraction simulators are exceptionally well suited for simulating short quantum circuits.
The simulation of Quantum Approximate Optimization Algorithm (QAOA) \cite{farhi2016quantum} circuits is exceptionally efficient with this approach given how short the circuits are.

In this work, we used our tensor network simulator QTensor, which is an open-source project developed in Argonne National Laboratory. The source code and documentation are available at \hyperlink{https://github.com/danlkv/QTensor}{github:danlkv/QTensor}.
It is a generic quantum circuits simulator capable of simulating generic quantum circuits and QAOA circuits in particular.
It also supports automatic differentiation with respect to gate parameters which is useful in the Quantum Machine Learning applications~\cite{Henry_ML} and QAOA parameter optimisation~\cite{Wurtz_angles}.

QAOA is a prime candidate to demonstrate the advantage of quantum computers in solving useful problems.
One major milestone in this direction is Google's simulations of random large quantum circuits \cite{arute2019quantum}.
The research interest of a community is now focused on providing an advantage of using quantum computers to solve real-world problems. QAOA is considered as a prime candidate to demonstrate such advantage.
QAOA can be used to solve a wide range of hard combinatorial problems with a plethora of real-life applications, 
like the MaxCut problem.
In this paper, we explored the limits of classical computing using a supercomputer to simulate large QAOA circuits, which in turn helps to define the requirements for a quantum computer to beat existing classical computers.

Our main contribution is the development of a novel slicing algorithm and
an ordering algorithm.
These improvements allowed us to increase the size of simulated circuits from 120
qubits to 210 qubits on a distributed computing system, while
maintaining the same time-to-solution.

In Section \ref{sec:related} we start the paper by discussing related work.
In Section \ref{sec:overview} we describe tensor networks and the bucket elimination algorithm.
Simulations of a single amplitude of QAOA ansatz state are described in Section \ref{sec:single_amp}.
We introduce a novel approach \stepslice to finding the slicing variables,
inspired by the tensor network structure.
Our algorithm allows simulating several amplitudes with little cost overhead, 
which is described in Section \ref{sec:multi_amp}.

We then show the experimental results of our algorithm running on 64-1,024 nodes of 
Argonne's Theta supercomputer.
All these results are described in Section \ref{sec:results}. In Section \ref{sec:conclusion} we  summarize our results and draw conclusions.

\section{Related Work}
\label{sec:related}

In recent years, much progress has been made in parallelizing state vector \cite{haner20170, smelyanskiy2016qhipster, wu2019full} and linear algebra simulators \cite{quac}. Very large quantum circuit simulations were performed on the most powerful supercomputers in the world, such as Summit \cite{villalonga2020establishing}, Cori \cite{haner20170}, Theta \cite{wu2019full}, and Sunway Taihulight~\cite{li2018quantum}.
All these simulators have various advantages and disadvantages. Some of them are general-purpose simulators, while others are more geared toward short-depth circuits.

One of the most promising types of simulators is based on the tensor network contraction technique.  This idea was introduced by Markov and Shi~\cite{markov2008simulating} and was later developed by Boixo et al.~\cite{Boixo2017} and other authors~\cite{SchutskiAdaptive}. Our simulator is based on representing quantum circuits as tensor networks.

Boixo et al.~\cite{Boixo2017} proposed using the line graphs of the classical tensor networks, an approach that has multiple benefits. 
First, it establishes the connection of quantum circuits with probabilistic graphical models, allowing knowledge transfer between the fields.
Second, these graphical models avoid the overhead of traditional diagrams for diagonal tensors.
Third, the treewidth is shown to be a universal measure of complexity for these models. It links the complexity of quantum states to the well-studied problems in graph theory, a topic we hope to explore in future works. Fourth, straightforward parallelization of the simulator is possible, as demonstrated in the work of Chen et al.~\cite{Chen2018}. The only disadvantage of the line graph approach is that it has limited usability to simulate subtensors of amplitudes, which was resolved in the work by Schutski et al. \cite{SchutskiAdaptive}.
%GAIL - well, if it was resolved, t is no longer a limitation.
The approach has been studied in numerous efficient parallel simulations relevant to this work \cite{Chen2018, li2018quantum, pednault2017breaking, SchutskiAdaptive}.
%GAIL - just a list doesn't really help unles you distinguish your work from theirs.

\section{Methodology}

\subsection{QAOA introduction}
\label{ssec:qaoa-intro}

The combinatorial optimization algorithms aim at solving a number of important problems. The solution is represented by an $N$-bit binary string $z=z_1{\dots z}_N$. The
goal is to determine a string that maximizes
a given classical objective function $C(z) : \{+1, -1\}^N$. The QAOA goal is to
find a string $z$ that achieves the desired approximation ratio:

\[ \frac{C(z)}{C_{max}} \geq r \]

where $C_{max}={max_z}C(z)$.

To solve such problems, QAOA was originally developed by Farhi et al.\cite{farhi2014quantum} in 2014. In this paper, QAOA has been applied to solve MaxCut problem. It was done by reformulating 
the classical objective function to quantum problem with replacing binary variables $z$ by quantum spin ${\sigma}^z$ resulting in the problem Hamiltonian ${H_C}$:

\[ {H_C=C({\sigma}^z_1, {\sigma}^z_2, {\dots}), {\sigma}^z_N} \]

After initialization of a quantum state $\ket{\psi_0}$, the ${H_C}$ and a mixing
Hamiltonian ${H_B}$:

\[ {H_B= \sum_{j=1}^N {\sigma}^j_x} \]

is then used as to evolve the initial state {p} times. It results in the variational wavefunction, which is parametirized by $2p$ variational parameters $\beta$ and $\gamma$. The ansatz state obtained after p layers of the QAOA is:

\[ {\ket{\psi_p(\beta,\gamma)} = \prod_{k=1}^{p} e^{-i\beta_p H_B} e^{-i\gamma_p H_C} \ket{\psi_0} }\]

To compute the best possible QAOA solution corresponding to the best objective function value, we need to sample the probability distribution of 
$2^N$ measurement outcomes in state $\ket{\gamma \beta}$. 
 %The need for sampling is resulting from the fact that the measurement outcomes can only take on discrete values in the form of an eigenstate.
 %To make it worse,
 The noise in actual quantum computers hinders the accuracy of sampling, resulting in the need of even a larger number of measurements.
 At the same time, sampling is an expensive process that needs to be controlled. Only a targeted subset of amplitudes
need to be computed because sampling all amplitudes will be very computationally expensive and memory footprint prohibitive. As a result, the ability of a simulator like QTensor to effectively sample certain amplitudes is a key advantage over other simulators.

The important conclusion Farhi et al.\cite{farhi2014quantum} paper was that to compute an expectation value, the complexity of the problem depends on the number of iterations $p$ rather than the size of the graph. This is a result of what is known as lightcone optimization. It has a major implication to the speed of a quantum simulator computing QAOA energy~\cite{Shaydulin2021_sym_acc}, but this type of optimizaiton is not applicable for simulating ansatz state, which is the type of simulation we focus in this paper. A more detailed MaxCut formulation for QAOA was provided by Wang et al.~\cite{wang2018quantum}. It is worth mentioning that there is a direct relationship between QAOA and adiabatic quantum computing, meaning that QAOA is a Trotterized adiabatic quantum algorithm. As a result, for large $p$ both approaches are the same.

\subsection{Description of quantum circuits}
\label{ssec:circuits}

A classical application of QAOA for benchmarking and code development is to apply it to Max-Cut problem for random 3-regular graphs.
A representative circuit for a single-depth QAOA circuit for a fully connected graph with 4 nodes, is shown in Fig. \ref{fig:circ}.
The generated circuit were converted to tensor networks as described in Section \ref{sec:circ_as_tensor}.
The resulting tensor network for the circuit in \ref{fig:circ} is shown in Fig. \ref{fig:graph_example}.
Every vertex corresponds to an index of a tensor of the quantum gate. Indices are labeled right to left: $0-3$ are indices of output statevector, and $32-25$ are indices of input statevector. Self-loop edges are not shown (in particular $Z^{2\gamma}$, which is diagonal).
\newcommand{\gbeta}{\vec \gamma, \vec \beta}
We simulated one amplitude of state $\ket{\gbeta}$ from the QAOA algorithm with depth $p=1$, which is
used to compute the energy function. 
The full energy function is defined by $\bra{\gbeta} \hat C \ket{\gbeta}$ and is 
essentially a duplicated tensor expression with a few additional gates from $\hat C$. 
The full energy computation corresponds to the simulation of a single amplitude of such duplicated tensor expression.

\section{Overview of simulation algorithm}
\label{sec:overview}

In this section, we briefly introduce the reader to the  tensor network contraction algorithm. It is described in much more detail in the paper by Boixo et al.  \cite{Boixo2017}, and the interested reader can refer to work by Detcher et al. \cite{detcher2013bucket} and Marsland et al. \cite{ marsland2011machine} to gain an understanding of this algorithm in the original context of probabilistic models.

% Theoretical description of garph model and bucket elimination

%%%%%%%%%%%
%%%%%%%%%%%

\subsection{Quantum circuit as tensor expression}
\label{sec:circ_as_tensor}

A quantum circuit is a set of gates that operate on qubits. 
% \todo{Describe what a subspace is}
% Not really essential
Each gate acts as a linear operator that is usually applied to a small 
subspace of the full space of states of the system.
State vector $\ket \psi$ of a system contains probability amplitudes for every possible configuration
of the system. A system that consists of $n$ two-state systems will have $2^n$ possible states
and is usually represented by a vector from $\mathbb C^{2^n}$.

However, when simulating action of local operators on large systems,
it is \add{more} useful to represent state as a tensor from $(\mathbb C^2)^{\otimes n}$
In tensor notation, an operator is represented as a 
tensor with input and output indices for each qubit it acts upon..
The input indices are equated with output indices of previous operator.
The resulting state is computed by summation over all joined indices. The comparison between Tensor Network notation and Dirac notation is shown in Table \ref{tab:notation}.

\begin{table}[b] \begin{tabular}{c|ll}
     &Dirac notation & Tensor notation \\
     \hline
     
     general&
     $\ket \phi = \hat X_0 \otimes \hat I_1 \ket \psi$ &
     $\phi_{i'j} = X_{i'i}\psi_{ij}$
     \\
     product state &
     $\ket \psi = \ket a \ket b$ & $\psi_{ij} = a_ia_j$
     \\
     with Bell state&
     $ \ket \phi = \hat X_0 \otimes \hat I_1 (\ket{00} + \ket{11}) $ &
     $ \phi_{i'j} = X_{i'i}\delta_{ij}$
\end{tabular}
\caption{Comparison between different notations of quantum circuits}
\label{tab:notation}
\end{table}

Following tensor notations we 
drop the summation sign over any repeated indices, that is, $a_i b_{ij} = \sum_i a_i b_{ij}$.
For more details on tensor expressions, see \cite{cichocki2016tensor}.

%\subsection{Tensor contraction ordering}
\subsection{Graph model of tensor expression}

\begin{figure}
       \definecolor{yellow}{rgb}{1.0, 0.76, 0.24}
    
    \begin{tabular}{cc}
    \begin{subfigure}{0.45\linewidth}
        \begin{subfigure}{0.25\linewidth}
            \Qcircuit @C=1em @R=.7em {
                & \ket{i} & & \gate{U} & \qw & \ket{i}
            }
        \end{subfigure}
        \hfill
        \begin{subfigure}{0.25\linewidth}
            \begin{tikzpicture}[scale=.7]
            %% vertices
            \fill[yellow] (0,0) circle (8pt);
            %% vertex labels
            \node at (0,0) {i};
            \end{tikzpicture}
        \end{subfigure}
        \vspace{2.7em}
        
        \begin{subfigure}{.25\linewidth}
                \Qcircuit @C=1em @R=.7em {
                    & \ket{i_1} & & \multigate{1}{U} & \qw & \ket{i_1}\\
                    & \ket{i_2} & & \ghost{U} & \qw & \ket{i_2}
                }
        \end{subfigure}
        \hfill
        \begin{subfigure}{.25\linewidth}
                \begin{tikzpicture}[scale=.7]
                \draw[thick] (0,0) -- (0,1);
                %% vertices
                \fill[yellow] (0,0) circle (8pt);
                \fill[yellow] (0,1) circle (8pt);
                \node at (0,0) {$i_1$};
                \node at (0,1) {$i_2$};
                %% vertex labels
                \end{tikzpicture}
        \end{subfigure}
        \vspace{0.4em}
        \caption{Diagonal gates}
    \end{subfigure}
    &
    \begin{subfigure}{0.5\linewidth}
        \begin{subfigure}{0.25\linewidth}
            \Qcircuit @C=1em @R=.7em {
                & \ket{i} & & \gate{U} & \qw & \ket{j}
                }
        \end{subfigure}        
        \hfill
        \begin{subfigure}{0.35\linewidth}
            \begin{tikzpicture}[scale=.7]
            \draw[thick] (0,0) -- (1,0);
            %% vertices
            \fill[yellow] (0,0) circle (8pt);
            \fill[yellow] (1,0) circle (8pt);
            \node at (0,0) {i};
            \node at (1,0) {j};
            %% vertex labels
            \end{tikzpicture}
        \end{subfigure}
        \vspace{2em}
    
        \begin{subfigure}{.20\linewidth}
            \Qcircuit @C=1em @R=.7em {%
                & \ket{i_1} & & \multigate{1}{U} & \qw & \ket{j_1}\\
                & \ket{i_2} & & \ghost{U} & \qw & \ket{j_2}
            }
        \end{subfigure}
        \hfill
        \begin{subfigure}{.4\linewidth}
            \begin{tikzpicture}[scale=.7]
            %%% edges
            \draw[thick] (0,0) -- (2,0) -- (1,0.5) -- (0,0) -- (1,1.5) -- (2,0) -- (1,0.5) -- (1,1.5);
            %% vertices
            \fill[yellow] (0,0) circle (7pt);
            \fill[yellow] (2,0) circle (7pt);
            \fill[yellow] (1,.5) circle (7pt);
            \fill[yellow] (1,1.5) circle (8pt);
            %% vertex labels
            \node at (0,0) {$i_1$};
            \node at (2,0) {$i_2$};
            \node at (1,0.5) {$j_1$};
            \node at (1,1.5) {$j_2$};
            \end{tikzpicture}
        \end{subfigure}
        \caption{Non-diagonal gates}
    \end{subfigure}
    \\
    \end{tabular} 
\caption{Correspondence of quantum gates and graphical representation.}
\label{fig:gadgets}
\end{figure}

\begin{figure}
  \centering
  \includegraphics[width=.9\linewidth]{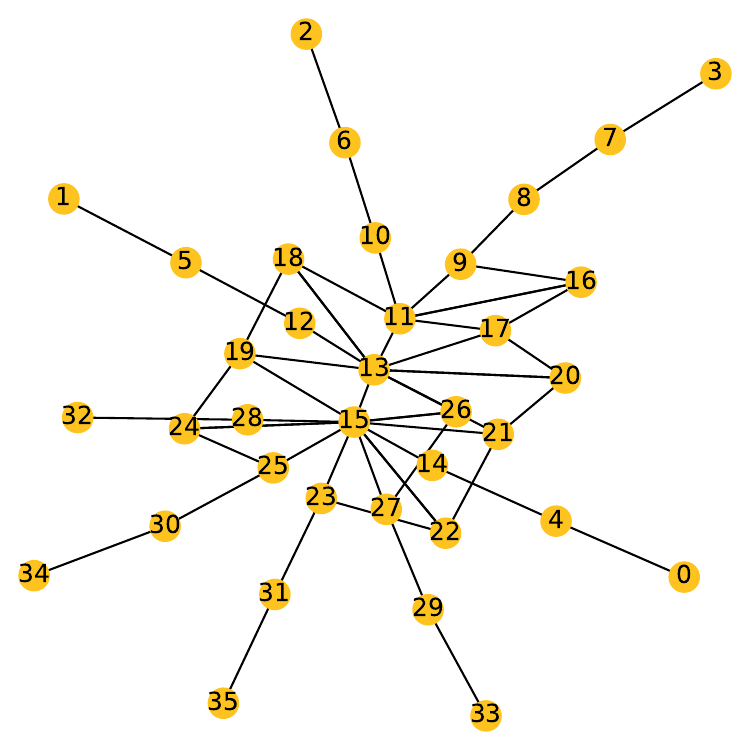}
  \caption{
  Graph representation of tensor expression of \add{the} circuit 
  in Fig. \ref{fig:circ}. 
  Every vertex corresponds to a tensor index of \add{a} quantum gate.
  Indices are labeled right to left: 0-3 are indices of \add{the} output statevector,
  and 32-25 are indices of \add{the} input statevector.
  Self-loop edges are not shown (in particular $Z^{2\gamma}$, which is diagonal).
  }
    \label{fig:graph_example}
\end{figure}

Evaluation of a tensor expression depends heavily on the 
order in which one picks indices to sum over \cite{schutski2019adaptive, markov2008simulating}.
The most widely used representation of a tensor expression is a ``tensor network,'', where vertices stand for tensors and tensor indices stand for edges.
For finding the best order of contraction for the expression, we use a 
line graph representation of a tensor network.
In this notation, we use vertices to denote unique indices, and we denote tensors by cliques (fully connected subgraphs).
Note that tensors, which are diagonal along some of the axes and 
hence can be indexed with fewer indices, are depicted by cliques that are smaller 
than the dimension of the corresponding tensor.
This results in a substantial simplification of the tensor network as described in detail in \cite{lykov_diagonal}.
For a special case of vectors or diagonal matrices, self-loop edges are used.
Figure~\ref{fig:gadgets} shows the notation for the gates used in this work.
For a more detailed description of graph representation, see \cite{schutski2019adaptive}.

Having built this representation, one has to determine the index elimination order.
The tensor network is contracted by sequential elimination of its indices. 

The tensor after each index elimination will be indexed by a union of sets of indices of tensors in the contraction operation.
In the line graph representation, the index contraction removes the corresponding vertices from the graph.
Adding the intermediate tensor afterwards corresponds to 
adding a clique to all neighbors of index $i$.
We call this step \emph{elimination of vertex (index) $i$}.
An interactive demo of this process can be found at 
%GAIL - this doesn't seem to comment out what you wish
%\%link to personal webapp, hidden for double-blind review, will be displayed in the final article\%
\url{https://lykov.tech/qg}
(works for cZ\_v2 circuits from ``Files to use''--- link).

\begin{figure}
  \centering
  \includegraphics[width=1.0\linewidth]{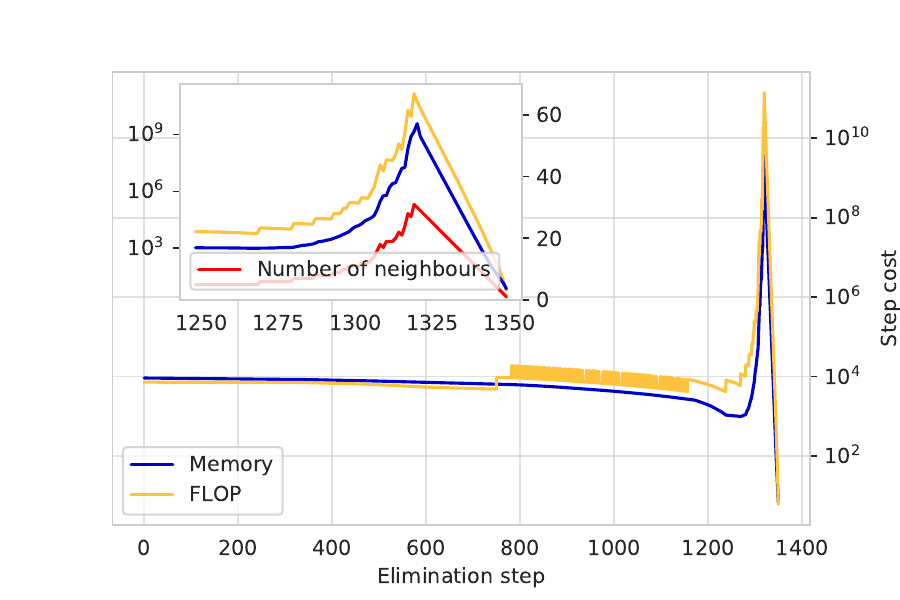}
  \caption{
  Cost of contraction \add{for} every vertex for a circuit with 150 qubits. 
  Inset shows the peak magnified and
  \add{the} number of neighbors of the vertex contracted at \add{a} given step (right y-axis).
  }
    \label{fig:contract_cost}
\end{figure}

The memory and time required for the new tensor after 
elimination of a vertex $v$ from $G$ depends exponentially on the number of its
neighbors $N_G(v)$.
Figure \ref{fig:contract_cost} shows the dependence of the elimination cost with respect to
the number of vertices (steps) of a typical QAOA quantum circuit. 
The inset also shows for comparison the number of neighbors for every vertex at the elimination step.

Note that the majority of contraction is very cheap, which corresponds to the low-degree
nodes from Figure \ref{fig:graph_example}.
This observation serves as a basis for our \stepslice algorithm.

The main factor that determines the computation cost is the maximum $N_G(v)$
throughout the process of sequential elimination of vertices. 
In other words, for the computation cost $C$ the following is true:
$$ C \propto 2^{\add{c}}; \add{c} \equiv \max_{i=1...N} N_{G_i}(v_i),$$
where $G_i$ is obtained by contracting $i-1$ vertices and $\add{c}$ 
is referred to as the \emph{contraction width}.
We later use  shorter notation for the number of neighbors $N_i(v) \equiv N_{G_i(v_i)}$.

The problem of finding a path of graph vertex elimination that minimizes $\add{c}$ is
connected to finding the tree decomposition.
In fact, the treewidth of the expression graph is equal to $\add{c}-1$.
Tree decomposition is NP-hard for general graphs~\cite{bodlaender1994tourist}, 
and a similar hardness result is known for the optimal tensor contraction problem~\cite{chi1997optimizing}.
However, several exact and approximate algorithms for tree decomposition were developed in graph theory literature;
for references,  see \cite{gogate2004complete, bodlaender2006exact, kloks1994treewidth, bodlaender1994tourist, kloks1993computing}.

%%%%%%%%%%%
%%%%%%%%%%%
    
 %  \begin{figure}
 %    \centering
 %    \includegraphics[width=0.9\linewidth]{figures/cost_vs_taskS_42d3.pdf}
 %    \caption{
 %            Theoretical estimation of resources required to simulate circuits \add{with different numbers of qubits}  without parallelization. Memory is plotted in amount of required complex numbers.
 %    }
 %    \label{fig:theor_cost}
 %  \end{figure}

\section{Simulation of a single amplitude}
\label{sec:single_amp}
    \newcommand{\zero}{\ket{0^{\otimes N}}}

The simulation of a single amplitude is a simple 
benchmark to use to evaluate the complexity of quantum circuits and simulation performance.
We start with $N$-qubit zero state $\zero$ and calculate a probability
to measure the same state.
$$
\sigma = \bra{0^{\otimes N}} \hat U \zero = \braket{0^{\otimes N} | \gbeta}
$$

\subsection{Ordering algorithm}
\label{sec:ordering}
\newcommand{\tw}{contraction width }
\newcommand{\rgr}{\emph{rgreedy} }

\begin{figure}
    \centering
    \includegraphics[width=.9\linewidth]{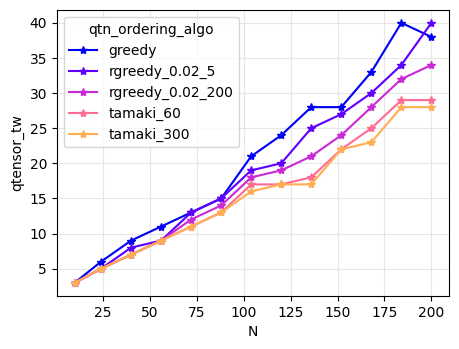}
    \caption{
    Comparison of different ordering algorithms for single amplitude simulation
    of QAOA ansatz state
    }
    \label{fig:ordering_algos}
\end{figure}

The ordering algorithm is a dominating part of efficient tensor network contraction.
Linear improvement in contraction width results in an exponential speedup of contraction.

 There are several ordering algorithms that we use in our simulations.
 The major criterion to choose one is to maintain a balance between ordering improvement 
 and run time of the algorithm itself.
 
\subsubsection{Greedy algorithm}

The greedy algorithm contracts the lowest-degree vertex in the graph.
This algorithm is commonly used as a baseline since it provides
a reasonable result given a short run-time budget.

\subsubsection{Randomized greedy algorithm}
The \tw is very sensitive to small changes in the contraction order.
Gray and Kourtis \cite{gray2020hyperoptimized} used this fact in a randomized ordering algorithm,
which provided \tw improvement without prolonging the run time.
We use a similar approach in the \rgr algorithm.
Instead of choosing the smallest-degree vertex, \rgr assigns probabilities for each vertex
using Boltzmann's distribution: 
$$p(v) = exp(-\frac{1}{\tau}N_G(v))$$

The contraction is then repeated $q$ times, and the best ordering is selected.
The $\tau$ and $q$ parameters are specified after the name of the \rgr algorithm.

\subsubsection{Heuristic solvers}

The attempt to use some global information in the ordering problem gives 
rise to several heuristic algorithms.

QuickBB \cite{gogate2004complete} is a widely-used branch-and-bound algorithm.
We found that it does not provide significant improvement in the \tw in addition to being much
slower than greedy algorithms.

Tamaki's heuristic solver \cite{tamaki:LIPIcs:2017:7880} is a dynamic programming approach
that provides great results.
This is also an ``anytime`` algorithm, meaning that it provides a solution after it is stopped at any time.
The improvements from this algorithm are noticeable when
it runs from tens of seconds to minutes.
We denote time (in seconds) allocated to this ordering algorithm after its name.

\section{Parallelization algorithm}
\label{sec:parallel}

\begin{figure*}
  \centering
  \includegraphics[width=0.9\textwidth]{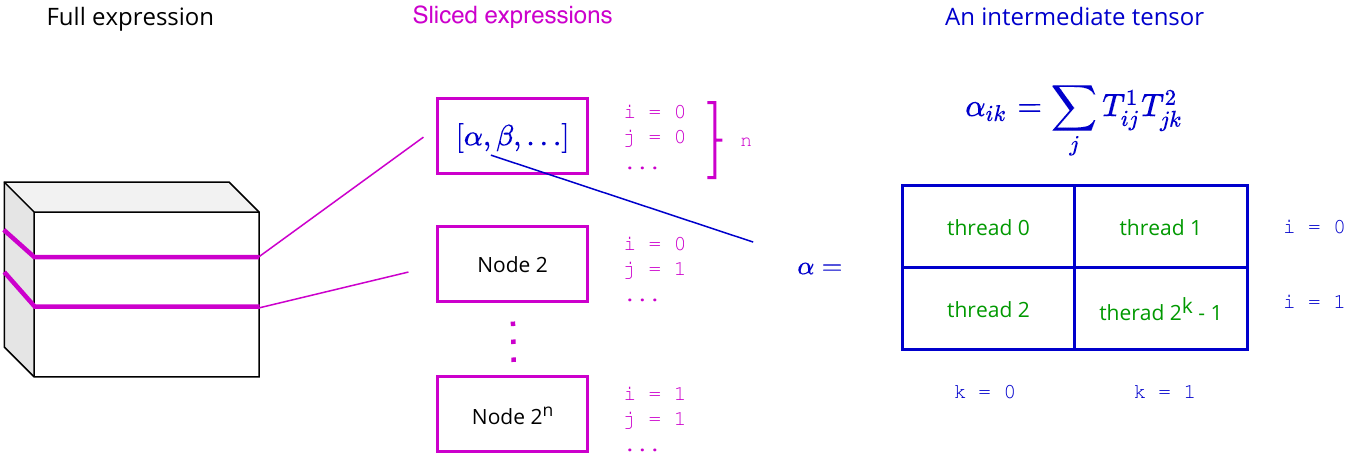}
  \caption{
  Illustration of our two-level tensor parallelization approach.
      On the multinode level MPI parallelization we use slicing of a partially contracted full expression.
      On the lower level of a single node, we use thread-based parallelization with a shared resulting tensor.
  }
  \label{fig:teaser}
\end{figure*}

We use a two-level parallelization architecture to couple the simulation structure and hardware constraints.
Our approach is shown at Fig. \ref{fig:teaser}.
Multinode-level parallelization uses MPI to share tasks. 
We slice the partially contracted full expression over $n$ indexes and distribute the slices to $2^n$ MPI ranks.
We use a novel algorithm for determining the slice vertex and step at which to perform slicing,
which results in massive expression simplification.
This is described in Section \ref{sec:mpi_par}. A high-level picture of our algorithm is shown in Fig. \ref{fig:flow_diagram}.

Node-level parallelisation over CPU cores uses system threads.
For every tensor multiplication and summation we slice the input and output tensors over $t$ indices.
Contraction is then performed by $2^t$ threads writing results to a shared result tensor.
This process is described in Section \ref{sec:single_node}.

To illustrate the two approaches used, we consider a simple expression \label{expr} $C_i = A_{ij}B_j$.
There are two obvious ways of parallelization:
\begin{enumerate}
    \item \label{par_opt1} Parallelization over elements of sum, index $j$. Every worker computes
    its version of $C_i$ for some value of $j$, and the results then are summed.
    \item \label{par_opt2} Parallelization over the indices of the result, $i$. 
    Every worker computes part of the result, $C_i$, for some value of $i$.
\end{enumerate}

These two options are intrinsically similar: every worker is assigned a simplified version of the expression, which is obtained by applying a slicing operation over some indices to every tensor. The difference between the two is that while performing computation using the first option, one must store copies of the result for every worker, which results in higher memory usage that scales linearly with the number of workers. This is not an issue in the second option, where different workers write to different parts of the shared result. The second option is less flexible, however. Usually one has a complex expression on the right-hand side, and the result has a smaller number of dimensions. 
The crucial part is that one can reduce treewidth of a complex expression using parallelization, which is discussed in  Section \ref{sec:mpi_par}.

\subsection{Description of hardware and software}
\label{ssec:supercomp}

The benchmarks reported in this paper were performed on the Intel Xeon Phi HPC systems in the Joint Laboratory for System Evaluation (JLSE) and the Theta supercomputer at the Argonne Leadership Computing Facility (ALCF) \cite{alcf}.
%which is a part of the U.S. Department of Energy (DOE) Office of Science (SC) Innovative and Novel Computational Impact on Theory and Experiment (INCITE) program \cite{incite}.
Theta is an 12-petaflop Cray XC40 supercomputer consisting of 4,392 Intel Xeon Phi 7230 processors. Hardware details for the JLSE and Theta HPC systems are shown in Table \ref{tab:hw}.

\begin{table}
  \caption{Hardware and software specifications}
  \label{tab:hw}
  %\begin{tabular}{p{0.37\columnwidth}p{0.55\columnwidth}}
  %\begin{tabular}{ll}
  \begin{tabularx}{\columnwidth}{X|X}
  \toprule
			\multicolumn{2}{c}{\textbf{\intelphi\ node characteristics}} \\
    \midrule 
    \intelphi\ models				&	7210 and 7230 (64~cores, 1.3~GHz, 
    									2,622 GFLOPs) \\
    Memory per node					&	16 GB MCDRAM, \newline 192 GB DDR4 RAM \\
    \midrule
    		\multicolumn{2}{c}{\textbf{JLSE \iphi\ cluster (26.2 TFLOPS peak)}} \\
    \midrule
    \# of \intelphi\ nodes	&	10 \\
    Interconnect type				&	Intel Omni-Path\textsuperscript{TM} \\
    \midrule
    		\multicolumn{2}{c}{\textbf{Theta supercomputer (11.69~PFLOPS peak)}} \\
    \midrule
    \# of \intelphi\ nodes				&		4,392 \\
    Interconnect type				&	Aries interconnect with \newline Dragonfly topology \\
    Cray environment loaded modules						&	PrgEnv-intel/ 6.0.5,
    intel/ 19.0.5.281, cray-mpich/ 7.7.10  \\
  \bottomrule
\end{tabularx}
%\end{tabular}
\end{table}

The Intel Xeon Phi processors used in this work have 64 cores. The cores operate at 1.3 GHz frequency. Besides the L1 and L2 caches, all the cores in the Intel Xeon Phi processors share 16 GBytes of MCDRAM (another name is High Bandwidth Memory) and 192 GBytes of DDR4 memory. The bandwidth of MCDRAM is approximately 400 GBytes/s, while the bandwidth of DDR4 is approximately 100 GBytes/s.

The memory on Xeon Phi processors can be configured in the following modes: flat mode, cache mode, and hybrid mode. In the flat mode, the two levels of memory are treated as separate entities. One can run entirely in MCDRAM or entirely in DDR4 memory. In the cache mode, the MCDRAM is treated as a direct-mapped L3 cache to the DDR4 layer. 
%At last, in the Hybrid mode the user is allowed to use a fraction of MCDRM as L3 cache and it allocates the rest of the MCDRAM as part of the DDR4 memory.
In the hybrid mode, a part of the MCDRAM is L3 cache and the rest is directly addressable fast MCDRAM, but it does not become part of the (lower bandwidth) DDR4 memory.

Besides memory modes, the Intel Xeon Phi processors support five cluster modes: all-to-all, quadrant/hemisphere, and sub-NUMA cluster SNC-4/SNC-2 modes of cache operation. The main idea behind these modes is how to optimally maintain cache coherency depending on data locality. 

For the types of problems we are computing here,  there is not much difference between various memory configurations \cite{Mironov2017sc}. In the calculations presented in this paper, we used the quadrant clustering mode for all quantum circuit simulations on Intel Xeon Phi nodes. 
We explored the use of different affinity modes and found that there is not much difference in performance between them. For our benchmarks, we used the default affinity, which is set to scatter.

\subsection{Single-node parallelization}
\label{sec:single_node}

%\iffalse
%\fi

Simulation of quantum circuits is an example of a memory-bound task: the main bottleneck of simulation is the storage of intermediate results of a simulation. In a simplistic approach called the state-vector evolution scheme, the full vector of size $2^n$ is stored in memory. Thus a circuit containing only 300 qubits will require more memory than there are atoms in the universe. 
A much more efficient algorithm is the tensor network contraction algorithm described here. But as we show below, it requires use of a complicated parallelization scheme compared with the straightforward linear algebra parallelization scheme used in the state-vector simulators.

Modern high-performance computing (HPC) systems have nodes with a large number of CPU cores.  An efficient calculation has to utilize all available CPU cores, using many threads to execute code. The major problem in using MPI-only code is that all of the data structures are replicated across MPI ranks, which results in increased memory usage linearly with respect to the number of MPI ranks. The largest data object in our simulation is the tensor, which is a result of the contraction step. Memory requirements to store such tensor are exponential with respect to its size.

Moreover, every code will inevitably have a part that can be executed only serially. As the number of OpenMP threads or MPI increases, the parallelization becomes less efficient according to Amdahl's law. Thus, following this logic, smaller computations require less time, and the portion of the program that benefits from parallelization will be smaller for small tensors. As a result, according to Amdahl's law, this means that for small tensors, we need to use  fewer threads.

To address these problems, we share the resulting tensor between $2^t$ threads.
We also use an adaptive thread count determined from task size (Eq. \ref{eq:adaptive_thread}).
A usual approach of splitting matrices in the code is to split into $2^t$ rows, or columns.
This approach is not applicable in our case since tensors have size 2 over each dimension, and it would require reshaping the tensor, so it would be indexed with a multi-index.
We choose a similar but more elegant approach.
To slice into $2^t$ parts, we first choose indices that will be our slice dimensions.
The slicing operation fixes the value of the index and reduces the number of dimensions by one.
We then use a binary form of the thread index (the id of the thread) as a point in space $\{0,1\}^{\otimes t}$ that defines the slice index values.

Every contraction in the bucket elimination step can be represented by the permutation of indices as

$$C_{ijk} = A_{ij}B_{ik}$$,

where index $i$ contains indices that $A$ and $B$ have in common
and $j,k$ contain indices specific to $A,B$, respectively.
For our simulation, we  slice the tensor over the first $t$ indexes of the resulting tensor because this approach results in consistent blocks of resulting tensors assigned to each thread, thereby
 reducing the memory access time.
%This approach also has perfect load-balance, since the size of tensors assigned to each thread are equal.
This part of the algorithm is shown in green in Fig \ref{fig:flow_diagram}.

To determine an optimal number of threads to use, we run a series of experiments to 
estimate the overhead time.
We use these experimental results as the basis for an empirical formula for optimal thread count:

    \begin{equation}
        t= \max(\lfloor \frac{r-22}{2} \rfloor, 1),
        \label{eq:adaptive_thread}
    \end{equation}
where $r$ is rank of the resulting tensor.

\subsection{Multinode parallelization}
\label{sec:mpi_par}

\begin{figure}
  \centering
  \includegraphics[width=0.9\linewidth]{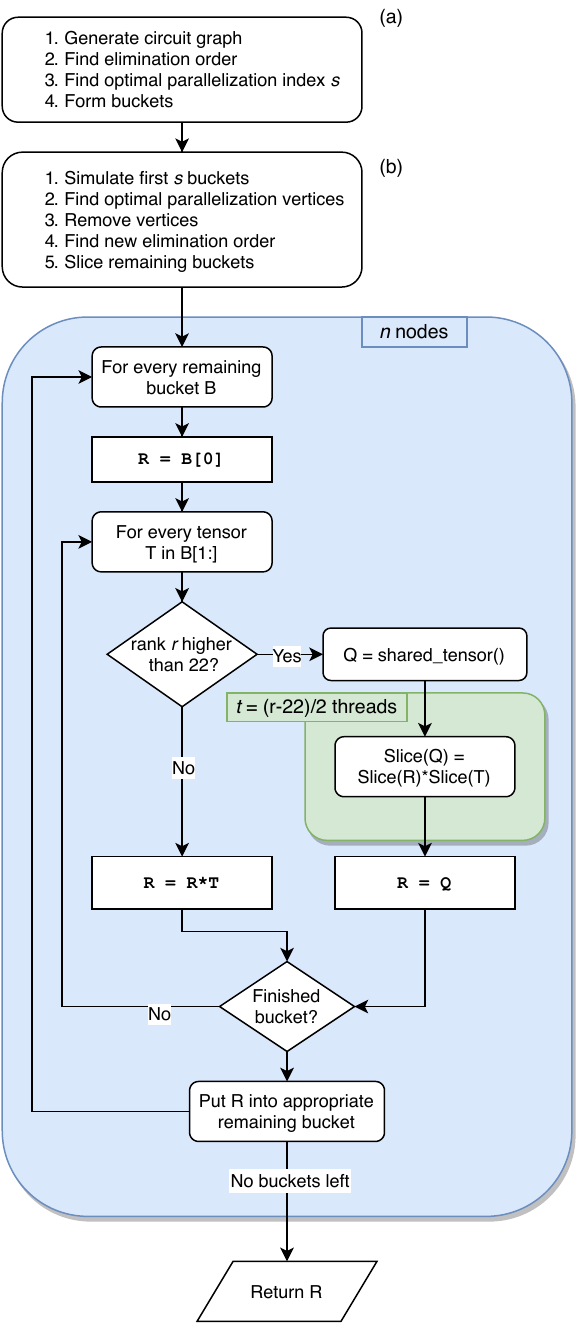}
  \caption{
    Sketch of the parallel bucket elimination algorithm.
    Part (a) and steps b2--b4 depend only on the structure of a task and can be executed only once for the QAOA algorithm.
    Steps b1 and b5 are performed serially.
   \\ 
    The outer loop of the blue region performs the elimination of the remaining buckets;
    the inner loop corresponds to processing a single bucket.
    The summation operation at the end of the bucket processing is omitted for simplicity.
  }
  \label{fig:flow_diagram}
\end{figure}

Every computational node has RAM and a pool of CPU cores. Parallelization over nodes (compared with threads) increases the size of aggregated distributed memory. Thus storing duplicates of tensors is not an issue. For this reason, we use every node to compute a version of a tensor expression evaluated at some values of the tensor indices.

In \emph{graph representation}, the contraction of the full expression is done by 
consecutive elimination of graph vertices. The elimination of a vertex removes
it from the graph and connects all neighbors. An interactive demo of this process can be found at 
%GAIL - this doesn't seem to comment out what you wish
\%link to personal webapp, hidden for double-blind review, will be displayed in the final article\%
%\url{https://lykov.tech/qg}
(works for cZ\_v2 circuits from ``Files to use''--- link).

The slice of a tensor over an index can be viewed as 
the function of many variables evaluated at some value of one variable
$$f(x_1, x_2, \dots x_n)|_{x_1 = a} = \tilde f(x_2,\dots x_n)$$, 
where variables can have integer values $v_i \in [0,d-1]$.
Slicing  reduces the number of indices of the tensor by one,
Moreover, in graph representation, this operation results in the removal of the corresponding vertex from the expression graph.
Since all sizes of indices we use are equal to 2,
removal of $n$ vertices allows us to split the expression into $2^n$ parts.

This operation is equivalent to decomposition of the full expression into the following form:
\begin{equation}
    \sum_{m_1\dots m_n} ( \sum_{V\backslash \{m_i\}} T^1T^2\dots T^N ),
\end{equation}
where $m_i$ are indices that we  slice over and the parts of the expression 
correspond to the expression in parentheses.

Each part is represented by a graph with lower connectivity than
the original one. This dramatically affects optimal elimination path and, respectively, the cost of contraction.
Depending on the expression, we observed that using only two computational nodes can allow for
speedups of an order of $2^5$.

The QAOA circuit tensor expression results in a graph that has many low-degree vertices, as demonstrated in Fig. \ref{fig:graph_example} for a small circuit. As can be seen in Fig. \ref{fig:contract_cost}, most contraction steps are computationally cheap, and connectivity of a graph is low.  Vertices can be removed at any step of contraction, giving  rise to a completely new problem of finding an optimal step for slicing the expression. We use a simple brute-force algorithm to determine the optimal step at which to perform parallelization. First, we find the ordering for the full graph and analyze the number of neighbors in the contraction path at each step. Any step after the step with the peak number of neighbors is out of consideration since our goal is to lower this peak, and we have to contract the initial expression before parallelizing. 
For every step of $K$ steps before the peak, we remove from the graph $n$ vertices with the biggest number of neighbors
 and rerun the ordering algorithm to determine the new contraction width.
The vertices could be any vertices in the graph, including ``free'' (nonrepeated) indices.
Since the removed vertices have the biggest number of neighbors, they usually index several tensors,
and the expression includes a sum over them.
We found that this new width can be lower than the original by more than $n$, providing freedom for massive reduction in the
contraction cost, as discussed in Section \ref{sec:results}.
Step $s$ at which the width is minimal is to be used in the main run of the simulation.

To the best of our knowledge, this approach of \emph{late parallelization} was never described in 
previous work of this field.

In the first part of the full simulation, 
labeled (a) in Figure \ref{fig:flow_diagram}, we read the circuit, create the expression graph, find the elimination order, and form buckets.
We also find the best parallelization step $s$ and the corresponding index used in the parallel bucket elimination.
The simulation starts with contracting the first $s$ buckets, which is computationally cheap. 
After this we have some other tensor expression network, which also is represented by a partially contracted graph.
This expression is conceptually no different from the one we started with; however, its graph representation has much higher connectivity.

The pseudo-code for the next stage, parallel bucket elimination, is listed in Algorithm \ref{code:parallel_algo}. We first select $n$ vertices with the most number of neighbors and use corresponding indices to slice the remaining expression over. To determine values for slices, we use the binary representation of the MPI rank of the current node. We find a new ordering for the sliced expression to identify a better elimination path with removed vertices taken into account. After reordering the sliced buckets, we run our bucket elimination algorithm with parallel tensor contraction. For every pair of tensors in the bucket, we determine the size of the resulting tensor as a union of the set of indices of both tensors. We then determine whether it is reasonable to use parallel contraction by checking that $t$ calculated by Eq. \ref{eq:adaptive_thread} is greater than 0. To run multiplication or summation in parallel, we first allocate a shared tensor, then perform the computation for slices of input and output tensors. The final result is obtained by summing the results from different nodes.

% Made algorithm more similar to on in detcher2013bucket
%\todo{fix accumulating contraction here}
\begin{algorithm}
\caption{Parallel bucket elimination}\label{code:parallel_algo}
\begin{algorithmic}[1]
\Require Ordered buckets $B_i$ containing tensors,
parallelization step $s$, number of parallel vertices $n$
vertex ordering $\pi: V \rightarrow \mathcal{N}, ~~ \pi = \{(v_{i}, i)\}_{i=1}^{|V|}$

\Ensure 
\Statex
\State contract\_first$(s, B_i)$     \Comment{Serial part: contract first $s$ buckets}
\For {$i = 0, n$} \Comment{Find best index to slice along}
    \State $p_i = \mathrm{max\_degree\_vertex}(G)$
    \State $\mathrm{remove\_vertex}(G, p_i)$
\EndFor
\State $ \vec v \gets \mathrm{binary\_repr}(\mathrm{mpi\_get\_rank}())$
\For {$j = 0, n$} \Comment{Slice the expression}
    \State $B_i \gets B_i|_{p_j=v_j}$
\EndFor
\For{$i = s,|V|$}
    \State $v \gets \pi^{-1}(i)$
    \State $R \gets B_i[0]$
    \For {$T \in B_i[1:]$} \Comment{Process next bucket}
        \State $r \gets | T.\mathrm{indexes} \cup R.\mathrm{indexes} |$
            \Comment{Determine resulting size}
        \State $t \gets floor(\frac{r - 22}{2})$
        \If {t>0} \Comment{Contract in thread pool}
            \State $Q \gets$ shared\_tensor($r$)
            \State $\vec w \gets \mathrm{binary\_repr}(get\_thread\_num())$
            \State $\vec k \gets \mathrm{indices\_of}(Q)[:t]$
            %\For {j=0,m} \Comment{Populate shared result tensor}
            \State $Q_{v\dots}|_{k_j = w_j} \gets (Q_{v\dots}T_{v\dots})|_{k_j = w_j} $
            %\EndFor
            %\For {j=0,m} \Comment{Sum the result tensor}
            \State $R \gets Q$  \Comment{$R$ now points to shared memory tensor}
        \Else
            \State $R \gets RT$
        \EndIf
    \EndFor
    \State $R \gets \sum_v R$ \Comment{Parallel sum can be implemented in 
    same fashion as contraction above} 
        
    \If{$R ~ \mathrm{is ~ scalar}$}
        \State $\mathrm{result} \gets \mathrm{result} \cdot R$
    \Else
        \State $k = \pi(w), ~w$ is the earliest index of $R$ w.r.t $\pi$
        \State $B_{k} \gets B_{k} \cup R$
    \EndIf
\EndFor
\State result $\gets \mathrm{mpi\_reduce\_sum}(\mathrm{result})$
\Comment{Gather the results}
\State \Return result
\end{algorithmic}
\end{algorithm}

\subsection{Step-dependent slicing}
\label{sec:stepslice}

\begin{figure}
    \centering
    \includegraphics[width=0.9\linewidth]{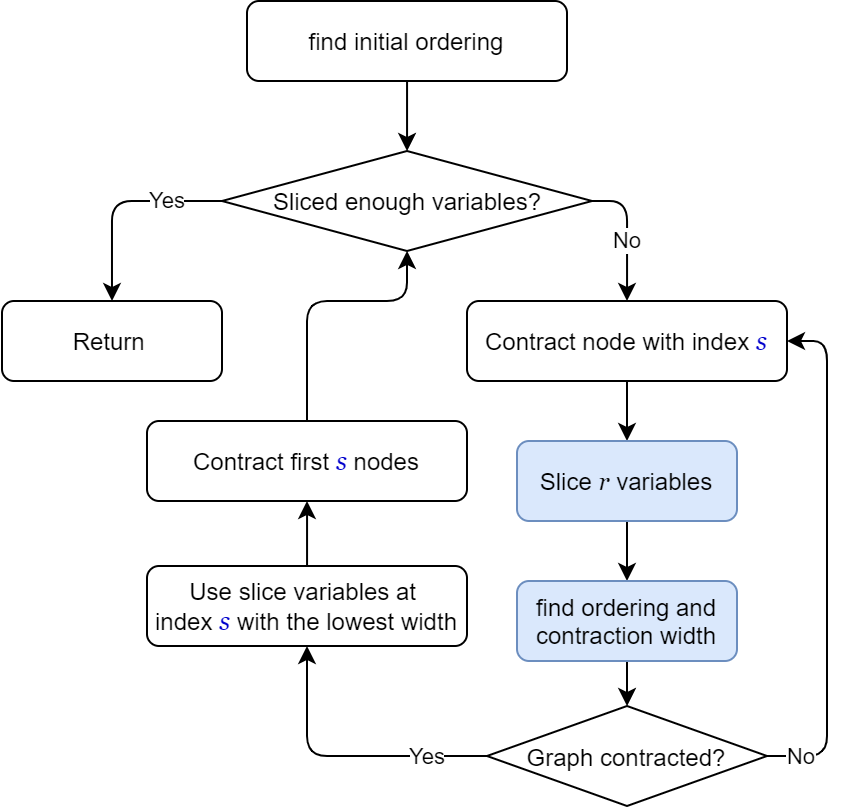}
    \caption{Step-based slicing algorithm. The blue boxes are evaluated for each graph node and are the main contributions to time.}
    \label{fig:slice_algo}
\end{figure}

The QAOA circuit tensor expression results in a graph that has many low-degree vertices, as
demonstrated in Fig. \ref{fig:graph_example} for a small circuit.
As can be seen in Fig. \ref{fig:contract_cost}, most contraction steps are computationally cheap,
and the connectivity of a graph is low. 

Each partially-contracted tensor network is a perfectly valid tensor network and can be 
sliced as well.
From a line graph representation perspective, vertices can be removed at any step of contraction,
giving  rise to a completely new problem of finding an optimal step for slicing the expression.
We propose a \stepslice algorithm that uses this fact and determines the best index to perform slicing operation, 
shown in Fig.~\ref{fig:slice_algo}.

We start with finding the ordering for the full graph.
 Our algorithm then selects consideration only those steps that come before the peak.
 For every such contraction step $s$, we remove $r$ vertices with the biggest number of neighbors from the graph and 
 re-run the ordering algorithm to determine the contraction after slicing.
 The distribution of \tw is shown on Fig.~\ref{fig:hist_of_cost}.
 
 The step $s$ at which slicing produces best \tw and contraction order before that is then added to 
 a contraction schedule.
 This process can be repeated several times until $n$ indices in total are selected - each $r$ of them having their optimal step $s$.
 This algorithm requires $\frac{n}{2r}\mathcal N$ runs of an ordering algorithm, where $\mathcal{N}$ is the number of nodes in the graph,
 which is usually of the order of 1000.
 Only greedy algorithms are used in this procedure due to its short run time.
 
 The value of $r$ can be used to slightly tweak the quality of the results.
 If $r=n$, all the $n$ variables are sliced at a single step.
 If $r=1$, each slice variable can have has its own slice step $s$, which gives better results for larger $n$.
 
We observed that using $n=1$ already provides \tw reduction by 3, which converts to 8x speedup in simulation.

To the best of our knowledge, this approach of \emph{step-dependent parallelization} was never described in 
previous work in this field.

\section{Simulation of several amplitudes}
\label{sec:multi_amp}

\begin{figure*}
    \centering
    \includegraphics[width=\textwidth]{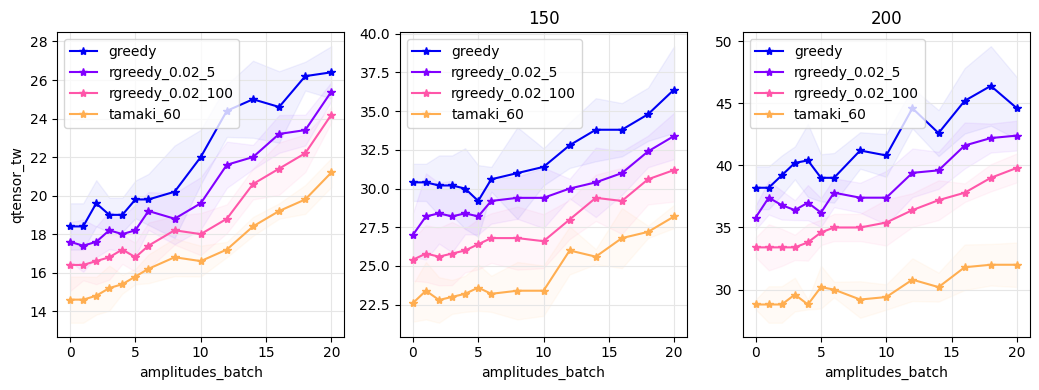}
    \caption{Simulation cost for a batch of amplitudes.
    The calculations are done for 5 random instances of degree-3 random regular graphs 
    and the mean value is plotted.
    The three plots are calculated for different number of qubits: 100, 150 and 200.
    }
    \label{fig:amplitude_batch}
\end{figure*}

The QAOA algorithm in its quantum part requires sampling of bit-strings that 
are potential solutions to a Max-Cut problem.
It is possible to emulate sampling on a classical computer without calculating all the 
probability amplitudes.
To obtain such samples, one can use \emph{frugal rejection sampling} \cite{villalonga2019flexible}
which requires calculating several amplitudes.

Our tensor network approach can be extended to simulate a batch of variables.
If we contract all indices of a tensor network, the result will be scalar - a probability amplitude.
If we decide to leave out some indices, the result will be a tensor indexed by those indices.

This tensor corresponds to a clique on left-out indices. 
If a graph contains a clique of size $a$, its treewidth is not smaller than $a$.
And if we found a contraction order with \tw $c$, during the contraction procedure we
will have a clique of size $c$.
If $a<c$ then adding a clique to the original graph does not increase \tw. 
This opens a possibility to simulate a batch of $2^a$ amplitudes for the same cost
as a single amplitude.
This is discussed in great detail in \cite{SchutskiAdaptive}.

Figure \ref{fig:amplitude_batch} shows \tw for simulation of batch of amplitudes for different values of $a$,
ordering algorithms and graph sizes.

\section{Results}
\label{sec:results}

\begin{figure}
  \centering
  \includegraphics[width=0.9\linewidth]{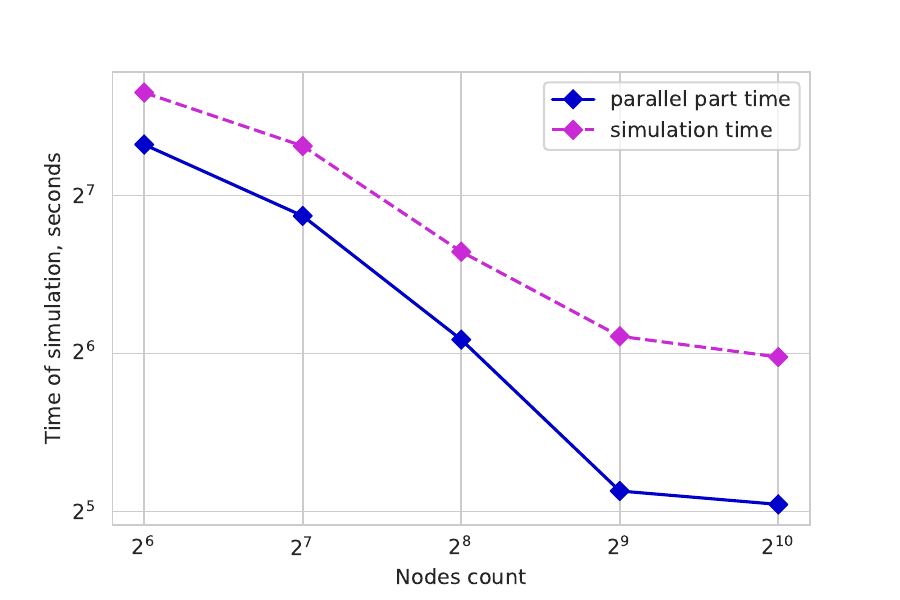}
  \caption{
      Experimental data of simulation time with respect to the number 
      of Theta nodes. The circuit is for 210 qubits and 1,785 gates.
  }
    \label{fig:full_sim}
\end{figure}

We used the Argonne's Cray XC40 supercomputer called Theta that consists of 4,392 computational nodes. Each node has 64 Intel Xeon Phi cores and 208 GB of RAM.
The combined computational power of this supercomputer is about 12 PFLOP/sec. The aggregated amount of RAM across nodes is approximately 900,000 GB.

 For our main test case, a circuit with 210 qubits, the initial contraction 
 was calculated using a greedy algorithm and resulted in \tw 44. This means that the cost of simulation would be $\geq$70 TFLOPS and 281 TB, respectively.
 Using our \stepslice algorithm with $r=n$ on 64 computational nodes allows us to remove 6 vertices and split the expression into smaller parts that have a contraction width of 32, which easily fits into RAM of one node. 
The whole simulation, in this case, uses 60\% of 13 TB cumulative memory of 64 nodes, more than 35x less than a serial approach uses.

\begin{figure}
  \centering
  \includegraphics[width=0.9\linewidth]{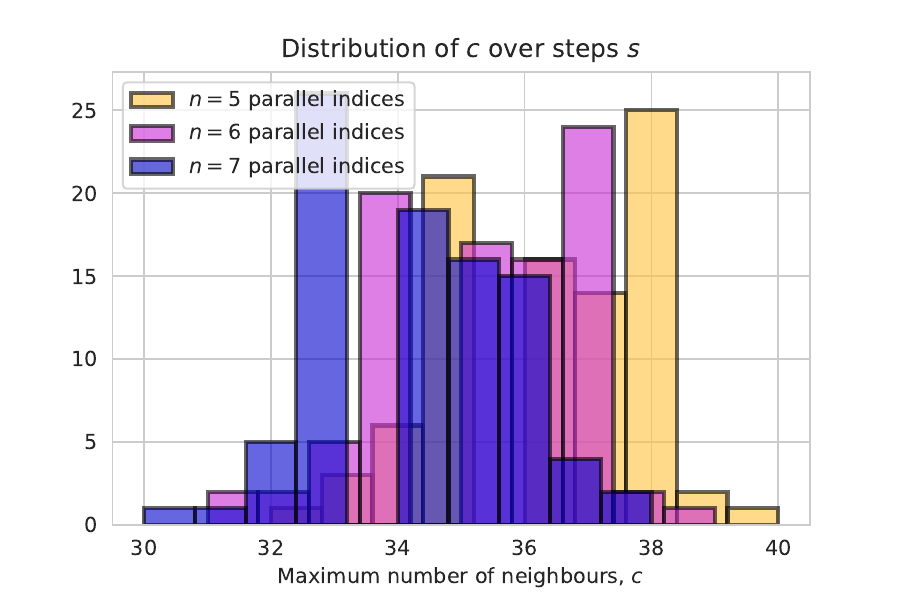}
  \caption{
    Distribution of the contraction width (maximum number of neighbors) $c$ for different numbers of 
    parallel indices $n$. While variance of $c$ is present, showing that it is sensible to 
    the parallelization index $s$, we are interested in the minimal value of $s$, which,
    in turn, generally gets smaller for bigger $n$.
  }
    \label{fig:hist_of_cost}
\end{figure}

Figure \ref{fig:hist_of_cost} shows how the contraction width $c$ of the sliced tensor expression depends on step $s$ for several
values of numbers of sliced indices $n$. 
The notable feature is the high variance of $c$ with respect to $s$---the difference between the smallest and the largest
values goes up to 9, which translates to a 512x cost difference. 
However, the general pattern for different QAOA circuits remains similar: increasing $n$ by one 
reduces $\min_s(c(s))$ by one.

Computational speedup provided by 64 nodes is on the order of $4096 = 2^{44-32}$ which is more than the theoretical limit of 64x 
for any kind of straightforward parallelization.
Using 512 nodes drops the contraction width to 29 and reduces the simulation time 3x compared with that when using 64 nodes. 

The experimental results for 64--1,024 nodes are shown in Fig. \ref{fig:full_sim}.
Simulation time includes serial simulation of the first small steps before step $s$,
which takes 40 s for a 210-qubit circuit, or 25--50\% of total simulation time, depending on the number of nodes.  

% Since simulation efficiency depends heavily on the contraction order defined by the expression structure,  situations exist in which doubling the number of nodes, while allowing  removal of an additional vertex, does not provide a reduction in contraction width. This is the case for our experiment when using 1,024 nodes resulting in the same contraction width of 29 as for 512 nodes. This behavior depends on the initial graph that we use for the Max-Cut problem. As a result, we do not expect speedup in such cases. This is the case for our circuit simulation shown in Fig. \ref{fig:full_sim}. The speedup for 1,024 vs. 512 nodes is not significant. Another notable effect comes from a different parallelization index $s$ for a different number of nodes. For 64 nodes, $s$ is bigger than for 128, which means that the parallelized expression for 128 nodes will contain more buckets. This can be observed in Fig. \ref{fig:chopping}. One can increase the speedup by careful management of tensors with different sizes. However, the main constraint for simulating quantum circuits is memory. In this work, we focused on increasing the maximum circuit size rather than maximizing speedup.

\section{Conclusions}
\label{sec:conclusion}

We have presented a novel approach for simulating large-scale quantum circuits represented by tensor network expressions. 
It allowed us to simulate large QAOA quantum circuits up to 210 qubit  circuits with a depth of 1,785 gates on 1,024 nodes and 213 TB of memory on the Theta supercomputer.

As a demonstration, We applied our algorithm to simulate quantum circuits for QAOA ansatz state with $p=1$, but our algorithm also works for higher $p$ also.
To reduce memory footprint, we developed a \stepslice algorithm that contracts part of an expression
in advance and reduces the expensive task of finding an elimination order.
Using this approach, we found an ordering that produces speedups up to 512x,
when compared with other parallelization steps $s$ for the same expression.

The unmodified tensor network contraction algorithm is able to simulate 120-140 qubit circuits, depending on the problem graph.
By using a randomized greedy ordering algorithm, we were able to raise this number to 175 qubits.
Furthermore, using a parallelization based on \stepslice allows us to simulate 210 qubits on the supercomputer Theta. 
Another way to obtain samples from the QAOA ansatz state is to use density matrix simulation, but it is prohibitively computationally expensive and memory demanding. The largest density matrix simulators known to us can compute 100 qubit problems \cite{Fried_2018qTorch} and 120 qubit problems \cite{zhao2020simulation} using high-performance computing.

The important feature of our algorithm is applicability to the QAOA algorithm: 
the contraction order has to be generated only once and then can be reused for additional simulations
with different circuit parameters. As a result, it can be used to simulate a large variety of QAOA circuits, which is used to study QAOA angles transferability between graphs \cite{Galda2021_transferab}, optimality of angles \cite{Wurtz_angles}, and the behaviour of QAOA under symmetry \cite{Shaydulin2021_symm}.

%In summary, the circuits we were able to simulate are much larger than the ones previously reported. Wu et al. \cite{wu2019full} were able to simulate quantum circuits up to 63 qubits using a lossy compression technique on the Theta supercomputer. Haner et al. \cite{haner20170} were able to simulate a maximum size of 45-qubit quantum circuits on the Cori supercomputer. We note that these simulators are state-vector simulators (meaning they simulate all amplitudes), whereas we simulate a single amplitude. Simulating a single amplitude using tensor networks offers unique advantages over state-vector simulators. Since we do not store full state vector, the tensor simulations are much more efficient by design. For circuits with high depth, however, state-vector simulators can outperform tensor network simulators. The direct comparison of our simulator with Intel-QS \cite{wu2019full} and ProjectQ \cite{haner20170} simulators is ambiguous. It will be the subject of our future work.

We conclude that this work presents a significant development in the field of quantum simulators. To the best of our knowledge, the presented results are the largest QAOA quantum circuit simulations reported to  date.

\section{Acknowledgements}
This research used the resources of the Argonne Leadership Computing Facility, which is a U.S. Department of Energy (DOE) Office of Science User Facility supported under Contract DE-AC02-06CH11357. We gratefully acknowledge the computing resources provided and operated by the Joint Laboratory for System Evaluation (JLSE) at Argonne National Laboratory.
This research was also supported by the U.S. Department of Energy, Office of Science, Basic Energy Sciences, Materials Sciences and Engineering Division, and by the Exascale Computing Project (17-SC-20-SC), a joint project of the U.S. Department of Energy’s Office of Science and National Nuclear Security Administration, responsible for delivering a capable exascale ecosystem, including software, applications, and hardware technology, to support the nation’s exascale computing imperative.

%\bibliographystyle{ACM-Reference-Format}
%\bibliography{sample-base}


\bibliographystyle{unsrt}  
\bibliography{qis}

\end{document}